\documentclass[fleqn,usenatbib]{mnras}

\usepackage[T1]{fontenc}

\DeclareRobustCommand{\VAN}[3]{#2}
\let\VANthebibliography\thebibliography
\def\thebibliography{\DeclareRobustCommand{\VAN}[3]{##3}\VANthebibliography}

\usepackage{graphicx}	
\usepackage{amsmath}	
\usepackage{amssymb}	
\usepackage[separate-uncertainty=true]{siunitx}
\usepackage{listings}
\usepackage{pdflscape}
\usepackage{soul} 

\newcommand{\code}[1]{\texorpdfstring{\textsc{#1}}{#1}}
\newcommand{\loge}{\TextOrMath{$\log_{\rm e}$}{\log_{\rm e}}}
\newcommand{\kms}{\TextOrMath{\,km~s$^{-1}$}{\mathrm{\,km~s}^{-1}}} 

\DeclareGraphicsExtensions{%
    .png,.PNG,%
    .pdf,.PDF,%
    .jpg,.mps,.jpeg,.jbig2,.jb2,.JPG,.JPEG,.JBIG2,.JB2}

\title[Constraining natal kicks]{New constraints on the Bray conservation-of-momentum natal kick model from multiple distinct observations}

\author[S. Richards et al.]{
S. M. Richards,$^{1}$\thanks{E-mail: sean.richards@auckland.ac.nz (SMR)}
J. J. Eldridge,$^{1}$
M. M. Briel,$^{1}$
H. F. Stevance$^{1,2}$
R. Willcox,$^{3,4}$
\\
$^{1}$Department of Physics, University of Auckland, Private Bag 92019, Auckland, New Zealand\\
$^{2}$Astrophysics Research Centre, School of Mathematics and Physics, Queen’s University Belfast, N. Ireland, BT7 1NN, United Kingdom.\\
$^{3}$School of Physics and Astronomy Monash University, Clayton, VIC 3800, Australia\\
$^{4}$The ARC Centre of Excellence for Gravitational Wave Discovery – OzGrav, Australia
}

\date{Accepted XXX. Received YYY; in original form ZZZ}

\pubyear{2022}

\usepackage{newtxtext,newtxmath}

\begin{document}
\label{firstpage}
\pagerange{\pageref{firstpage}--\pageref{lastpage}}
\maketitle

\begin{abstract}
Natal supernova kicks, the linear momentum compact remnants receive during their formation, are an essential part of binary population synthesis (BPS) models. Although these kicks are well-supported by evidence, their underlying distributions and incorporation into BPS models is uncertain. In this work, we investigate the nature of natal kicks using a previously proposed analytical prescription where the strength of the kick is given by $v_\text{k}=\alpha\frac{m_\text{ejecta}}{m_\text{remnant}}+\beta \kms$, for free parameters $\alpha$ and $\beta$. We vary the free parameters over large ranges of possible values, comparing these synthetic populations simultaneously against four constraints: the merger rate of compact binary neutron star (BNS) systems, the period-eccentricity distribution of galactic BNSs, the velocity distribution of single-star pulsars, and the likelihood for low-ejecta mass supernovae to produce low-velocity kicks. We find that different samples of the parameter space satisfy each tests, and only 1 per cent of the models satisfy all four constraints simultaneously. Although we cannot identify a single best kick model, we report $\alpha=115^{+40}_{-55}\kms, \beta=15^{+10}_{-15}\kms$ as the center of the region of the parameter space that fulfils all of our constraints, and expect $\beta\geq0\kms$ as a further constraint. We also suggest further observations that will enable future refinement of the kick model. A sensitive test for the kick model will be the redshift evolution of the BNS merger rate since this is effectively a direct measure of the delay-time distribution for mergers. For our best fitting values, we find that the peak of the BNS merger rate is the present-day.
\end{abstract}



\section{Introduction}

At the terminus of a massive star's ($M_\text{initial} \gtrsim 8M_\odot$) lifetime, it experiences a core-collapse supernova (CCSNe). This supernova imparts velocity onto the remnant it leaves behind - termed a natal supernova kick. Although evidence for a natal kick has been discussed since at least the 1960s \citep[e.g.,][]{Blaauw_1961, Ostriker1970Pulsars, deLoore_1975, Sutantyo_1978}, there are still two areas of active research. First, how the kick is caused in the core-collapse. Although it is generally understood to be the result of conservation of momentum, it is uncertain whether this is caused by an asymmetric mass ejection during the supernova or anisotropic emission of neutrinos during the cooling of the compact remnant \citep[][and references therein]{Wongwathanarat_2013_tugboat}. Secondly, and of particular interest to our work, is the distribution of velocities imparted on the remnants \citep[e.g.,][]{Hobbs2005AMotions, Baker_2008, Wongwathanarat_2013_tugboat, Bray2018_RefiningKick, Giacobbo_2020, mandelmuller}. The velocity distribution provided by \citet{Hobbs2005AMotions} is generally used, however there are conflicting accounts \citep[cf.][]{Fryer_1998, Arzoumanian_2002, Bombaci_2004, VerbuntIgoshevCator} of the precise nature of the distribution.

These conflicting accounts, in part, stem from the two approaches to prescribing a kick distribution. These are based in empirically inferential methods and theoretical prescriptions. Empirically inferential methods here mean statistical fits to observed pulsar velocities, such as the fit to runaway pulsars by \citet{Hobbs2005AMotions}, whereas a theoretical prescription defines the kick velocity in terms of physical parameters of the supernova or star system, such as the ejecta or remnant masses. An example of this prescription is equation 2 of \citet{mandelmuller}, which takes the neutron star kick velocity as proportional to the CO core mass and the neutron star mass. Both approaches have their merits. The empirical approach can include observational data that challenges theoretical understanding. On the other hand, theoretical prescriptions allow us to make predictions which can be validated against future observations. However, both approaches suffer from drawbacks: empirically inferential methods are subject to biases in their selection. Conversely, theoretical prescriptions are ultimately a parameterisation of complex physical phenomena. They can suffer from simplifications or assumptions that introduce difficult to quantify inaccuracies.

Two kick prescriptions we examine in detail in this work are the theoretically prescribed kick from \citet{Bray2016_ProposingKick, Bray2018_RefiningKick} (hereinafter the `Bray kick') and the empirically inferred kick from \citet{Hobbs2005AMotions} (the `Hobbs kick'). The reason we select these two kicks is twofold. We use the Hobbs kick as our reference kick as it has been widely traversed in literature \citep[see, among others,][]{Nakar_2007, Belczynski_2008, Lorimer_2008, Dominik_2012, Eldridge_2017}, and the Bray kick as its two free parameters, $\alpha$ and $\beta$, let us generate a population of kicks tied to the properties of the progenitors. We also consider the Bray kick and the Hobbs kick as they have differing physics underlying their selection of the kick mechanism. The Bray kick originates in an argument from conservation of momentum \citep[as argued in][]{Janka_2017}, whereas the Hobbs kick is a random kick sampled from a Maxwell-Boltzmann distribution centered on $\sigma=265\kms$, chosen as an empirical fit to the proper motions of 233 pulsars. Both kicks sample the direction isotropically. The Bray kick takes the kick velocity to be linear in the ejecta-remnant mass ratio,

\begin{equation}
    v_\text{k} = \alpha\left(\frac{M_\text{ejecta}}{M_\text{remnant}}\right)+\beta\left(\frac{M_\text{NS}}{M_\text{remnant}}\right),
\end{equation}

with the most recent fiducial parameters $\alpha = 100^{+30}_{-20}~\si{\km\per\second}$ and $\beta = -170^{+100}_{-100}~\si{\km\per\second}$ from \citet{Bray2018_RefiningKick}, and $M_\text{NS}$ the mass of a neutron star. We note that the Bray kick has an extra $M_\text{NS} / M_\text{rem}$ as a coefficient to $\beta$, as in \citet{Ghodla_2021}. This factor was not present in the original kick proposed by \citet{Bray2016_ProposingKick}, and has been introduced in work since then. The reason for this inclusion is it allows the same kick model to be extrapolated for use with the natal kicks received by more massive black holes. We assume that for supernovae resulting in a black hole, $M_\text{NS}=1.4M_\odot$, and for neutron stars (as in this work), the coefficient reduces to unity as the remnant mass is, by definition, the mass of a neutron star. On the other hand, the Hobbs kick samples the velocity from a Maxwell-Boltzmann distribution, given by

\begin{equation}
    P(v_\text{k}){\rm~d}v_\text{k} = \sqrt{\frac{2}{\pi}}\frac{v_\text{k}^2}{\sigma^3}{\rm exp}\left(-\frac{v_\text{k}^2}{2\sigma^2}\right){\rm~d}v_\text{k},
    \label{eq:hobbskick}
\end{equation}

with the distribution parameter fixed at $\sigma = \SI{265}{\km\per\second}$.

Some authors allow the distribution parameter of the Hobbs kick, $\sigma$, to vary, whilst some take a bimodal approach \citep[see, among others,][]{Arzoumanian_2002, Vigna-Gomez2018OnStars, Beniamini_2016}. On the other hand, the Bray kick can take a wide range of values in its two free parameters, $\alpha$ and $\beta$, which we assume to be constant for all supernovae.

In this work we do not consider a bimodal Hobbs kick, such as those described in \citet{Arzoumanian_2002, Bombaci_2004, VerbuntIgoshevCator}. Bimodal kicks typically take a second peak near $v_\text{k}\simeq30\kms$, to provide peaks for ultra-stripped supernovae (USSNe) and electron-capture supernovae (ECSNe). Both USSNe and ECSNe are often associated with weak kicks \citep{Willcox_2021} -- in the case of USSNe, likely due to the low ejecta mass, and in the case of ECSNe due to the relative difficulty of forming anisotropies in the supernova \citep{GiacobboMapelli_2018}. Conversely, kicks on the order of ${\rm few}\times 100\kms$ are associated with CCSNe, and likely result from small asymmetries in the supernova providing an asymmetric mass ejection \citep{LyneLorimer_1994, Wongwathanarat_2011}.

\begin{table}
    \centering
    \begin{tabular}{c|c|c|c}
        \hline
        Catalogue & Model & $\mathcal{R}_0$ / \# yr$^{-1}$ Gpc$^{-3}$ & Ref. \\
        \hline
        GWTC-2 & - & $320^{+490}_{-240}$ & $^*$ \\
        GWTC-3 & BGP & $99^{+260}_{-86}$ & $^\dag$ \\
        GWTC-3 & MS & $470^{+1430}_{-413}$ & $^\dag$ \\
        GWTC-3 & PGB (ind) & $250^{+640}_{-196}$ & $^\dag$ \\
        Theoretical & - & $407^{+19}_{-19}$ & $^\ddag$ \\
        \hline
    \end{tabular}
    \caption[]{The merger rates for BNS systems, as reported by the LVK. In GWTC-3, we do not report their PGB (pair) rate, as that does not contribute to their union of credible intervals and therefore to the overall reported merger rate. Acronyms are taken from their sources: PDB is the \textsc{Power Law + Dip + Break} model (with no pairing function), MS is the \textsc{Multi source} model, and BGP is the \textsc{Binned Gaussian process} model.
    
    \textbf{Sources:}
    
    $^*$ \citet{GWTC-2:Data}
    
    $^\dag$ \citet{GWTC-3:Data}
    
    $^\ddag$ \citet{EldridgeStanwayTang_2018}}
    \label{tab:ligo_rates}
\end{table}

In this paper we constrain the kick distribution by using binary population synthesis (BPS) models and comparing these synthetic populations to multiple observations in order to determine if a single kick can explain all our observations simultaneously. These observations are:

\begin{enumerate}
    \item Gravitational wave (GW) transient rates by the LIGO-Virgo-KAGRA collaboration (\citet{GWTC-2:Data, GWTC-3:Data}, hereafter LVK),
    \item Observations of galactic BNS systems \citep[][and references therein, summarised \autoref{tab:observed_BNSes}]{Vigna-Gomez2018OnStars},
    \item The velocity distributions of single-star pulsars \citep[][and references therein]{Willcox_2021}, and
    \item The kick velocities of USSNe.
\end{enumerate}

Previous work on narrowing down the nature of the natal kick such as \citet{Bray2018_RefiningKick} has focused on a subset of these observational constraints. However, with the benefit of larger and more precise datasets, we are now in a position to apply multiple constraints to our models. Using multiple observational datasets allows us to create orthogonal, independent, constraints on our values of $\alpha$ and $\beta$.

It is pertinent to note that we use the terminology `Binary Neutron Star' (BNS) instead of `Double Neutron Star' (DNS): DNS systems include binary systems formed by, for example, the capture of a runaway star by an isolated star. On the other hand, BNS systems are the subset of DNS systems which are formed as binaries from birth. Our analysis is focussed on BNS systems.

This paper is structured as follows: in Section~\ref{sec:methodology}, we outline our methodology for the BPS of, and inspiral calculations for, compact BNS systems, including the BPASS project, our code base, and the statistical methods we utilise. In Section~\ref{sec:results}, we examine the results of these statistical tests, and in Section~\ref{sec:discussion}, we discuss our results in the broader picture of BNS evolution.

\section{Methodology}\label{sec:methodology}

\subsection{The BPASS project}\label{sec:datagen}

Our BPS is performed using the Binary Population and Spectral Synthesis (BPASS) suite of codes, described by \citet{Eldridge_2017, Stanway_2018}. We use the fiducial version 2.2.1 models, which implement the initial mass function (IMF) from \citet{Kroupa_IMF}, take an upper mass limit of $300 M_\odot$, and use the initial binary parameters of \citet{MoeDiStefano_2017}.

The BPASS models are a set of binary and single-star models generated through a modified version of Cambridge \code{stars} code (originally described in \citet{Eggleton_1971} and most recently described in \citet{Eldridge_2017}), with BPS performed by the \textsc{Tui} code. \code{Tui} uses the input stellar population from the fiducial BPASS models and follows the evolution of the binary until the secondary supernova. It records the final BNS population and estimates the time the binary will take to merge due to the emission of gravitational radiation. \code{Tui} further extracts the velocity distribution of isolated pulsars.

Our models follow the input file format described in \citet{Eldridge_2017}, pairing primary and secondary model files. Briefly, to use compuational resources efficiently, BPASS only calculates the detailed structure and evolution of one star at a time. The most massive star, the primary, is evolved in detail first while the secondary's evolution is approximated by using the equations of \citet{Hurley_2002}. These are referred to as our {\em primary models}.

After the first supernova, the evolution of the secondary is recomputed using a detailed evolution model. If accretion onto the secondary happened during the primary evolution then the maximum mass of the secondary is used. If in the first supernova the system is unbound then a single star is used. However, when the system remains bound, the secondary star is evolved in detail, modeling the compact remnant as a white dwarf, neutron star or black hole, depending on its mass. These are our {\em secondary models}. The initial period used for this secondary binary model is given from analysing the effect of the kick on the pre-SN orbit. 

Then, for each kick model -- either a pair of $\alpha$ and $\beta$ parameters, or the Hobbs kick -- we simulate 1,000 kicks. This results in a selection of systems which have the same masses, with different eccentricities, periods, and ages, as well as kick velocities.

In this paper, we only consider solar metallicity (which we take to be $Z_\odot \equiv 0.020$), as \citet{Tang_2020} showed that there is only a weak influence on BNS merger rate densities by the metallicity for BNS systems. We expect the other tests to also be only weakly dependent on metallicity, as we are using young BNS systems recently formed, and the BNS distribution is linked to the BNS merger rate, indicating weak dependence.

We have created three different sets of \code{Tui} populations in this work. These are:

\begin{enumerate}
    \item A fiducial population based on \citet{Hobbs2005AMotions},
    \item A wide grid of populations with $\alpha\in\left[0,700\right] \kms$ and $\beta\in\left[-350,350\right] \kms$ with a step size in both directions of 50 $\kms$ (\textsc{WideGrid}), and
    \item A fine grid of populations with $\alpha\in\left[0, 400\right]\kms$ and $\beta\in\left[-200, 200\right]\kms$ with a step size in both directions of 10 $\kms$ (\textsc{FineGrid}).
\end{enumerate}

Both \textsc{FineGrid} and \textsc{WideGrid} are computed as isolated binary grids and separately as distributions of isolated single-star pulsar velocities.

\subsection{\code{Takahe}: Modelling compact remnant binary populations and their mergers}

\code{Tui} only provides the BNS populations at the moment after the second supernova, and thus we must compute their coalescence time. We use the equations formulated by \citet{Peters1964GravitationalMasses} to simulate the inspiral of compact remnants. These are

\begin{align}
    \frac{\mathrm{d}a}{\mathrm{d}t} &= \frac{-\beta}{a^3\left(1-e^2\right)^\frac72}\left(1+\frac{73}{24}e^2+\frac{37}{96}e^4\right) \label{eq:dadt}\\
    &\text{and}\notag\\
    \frac{\mathrm{d}e}{\mathrm{d}t} &= -\frac{19}{12}\frac{\beta}{a^4\left(1-e^2\right)^\frac52}\left(e+\frac{121}{304}e^3\right). \label{eq:dedt}
\end{align}

It is pertinent to note that these equations are formulated as the equations of motion in the weak field regime. They do not account for general relativistic effects past the innermost stable circular orbit $r_\text{ISCO}=\frac{6GM}{c^2}$ of the primary star. However, based on prior work with the \textsc{riroriro} code \citep{vanZeist_2021}, which accounts for up to third post-Newtonian terms, we expect that the merger time-scale will only be higher by fractions of years in this regime, and thus we do not consider effects stronger than this.

There are no closed-form solutions to these coupled, inhomogenous, differential equations, although approximations do exist, such as \citet{mandel2021accurate}. It can be shown that the coalescence time of the binary system, given initial semimajor axis $a_0$ and eccentricity $e_0$ is given by

\begin{equation}
    T_\text{c}(a_0, e_0) = \frac{12}{19}\frac{c_0^4}{\beta}\int_0^{e_0} \frac{e^\frac{22}{19}\left(1+\frac{121}{304}e^2\right)^\frac{1181}{2299}}{\left(1-e^2\right)^\frac32}{ \mathrm{d}e},
\end{equation}

with $c_0$ a constant, described in \citet{Peters1964GravitationalMasses}. We therefore take the total merger time as

\begin{equation}
    \tau = T_\text{evo} + T_\text{rejuv} + T_\text{c}.
\end{equation}

Here $T_\text{evo}$ is the evolution age -- the length of time from the secondary's zero-age main sequence until its death. $T_\text{rejuv}$ is the stellar rejuvenation age. This factor is included to account for the effects of rejuvenation of the secondary star in a binary, which is only included if during the primary model the secondary's star's mass increases due to mass transfer. If the companion star accretes mass, we assume it has rejuvenated to the zero-age main sequence (ZAMS). We record the time when the companion star last accreted material, and if it accreted more than 5 per cent of its initial mass, then we add this age onto the total merger time. The BPASS codes assume that the accretor is always a main sequence star, see the end of Section 7.1 in \citet{Eldridge_2017} for further details on the rejuvenation age treatment.

The four parameters, $m_1, m_2, a, e$, taken at time $t=0$, determine the future evolution of the compact binary remnant, and thus we have a four-dimensional phase space. However, as the system contains two compact objects, we treat the masses as constant throughout the evolution, and thus there are only two time-varying parameters: $a$ and $e$.

We use the \code{Takahe} library to solve the coupled equations Equation~\ref{eq:dadt} and Equation~\ref{eq:dedt} for each set of parameters within the parameter spaces defined in Section~\ref{sec:datagen}. \code{Takahe} is a \code{Python} \citep{python} library with a \code{Julia} \citep{bezanson2014julia} integrator, that allows us to efficiently solve Equations \ref{eq:dadt} and \ref{eq:dedt}.

\code{Takahe} imposes both a maximum cut-off time for the integration of $10^{11}$ years, as well as a minimum semi-major axis of 10 km. The former is chosen as a value past the age of the Universe, whilst the latter is chosen to be the average radius of a neutron star \citep[$\sim12$ km, per][]{Lattimer_2015}. This ensures that the integrator will either evolve until coalescence, or until the system has lived too long to be observed. \code{Takahe} also assumes a $\Lambda$CDM cosmology, with $\Omega_\text{M}=0.3$, $\Omega_\Lambda=0.7$, and $H_0=70 \kms~{\rm Mpc^{-1}}$ for the purposes of converting between redshift and lookback time. \code{Takahe} is available on GitHub, see the \hyperref[sec:dataavailability]{Data Availability Statement} for details.

Our stellar formation rates are taken from \citet{Madau2014CosmicHistory},

\begin{equation}
    \psi(z) = 0.015\frac{(1+z)^{2.7}}{1+((1+z)/2.9)^{5.6}}~\text{M}_\odot~\text{yr}^{-1}~\text{Mpc}^{-3}.
\end{equation}

The BNS merger rate density, $\mathcal{R}$, as a function of redshift is therefore given by the convolution

\begin{equation}
    \mathcal{R}(z) = \psi\left(z\right) * m(t_z),
\end{equation}

where $m(t_z)$ is the content of the delay-time distribution (DTD) at lookback time $t_z$ corresponding to the redshift $z$. We denote the BNS merger rate we detect today as $\mathcal{R}_0 \equiv \mathcal{R}(0)$. Computationally, we use the binned histogram method described in in Section 4 of \citet{Briel_2022}, with the caveat that, since our calculation is only performed for solar metallicity, our SFH does not incorporate the cosmic metallicity distribution from \citet{Langer2006ONEVOLUTION, EldridgeStanwayTang_2018, Tang_2020}

In this project, we only consider neutron stars, and assume that those have masses $M_\text{rem}\leq 2.5M_\odot$. Recent studies such as \citet{Linares_2018} have discovered neutron stars with masses $m=2.17_{-0.15}^{+0.17}$, so our threshold admits those systems. For further information on the selection of the upper bound, the reader is referred to Section 3.2 of \citet{Stevance_SLSN}.

\subsection{Statistical Methods}

\subsubsection{BNS merger rates: The O3a Contour}\label{sec:mergerrates}

Our first approach is to examine the merger rate today of BNS systems, $\mathcal{R}_0$. We evolve every model in our dataset, predict the BNS merger rate at $z=0$, and use this to compute the DTD. We then extract all combinations of $\alpha$ and $\beta$ that result in a value of $\mathcal{R}_0$ within the contour defined by the LIGO-Virgo O3a merger rate.


\subsubsection{Period-Eccentricity distributions: Bayes Factors}\label{sec:bayesfactors}

We quantify the goodness-of-fit of the period-eccentricity distributions by computing a probability map. Firstly, we divide the space into 2D bins, which have widths of 0.05 in eccentricity space, and 0.2 dex in $\log~P$ space. For each system, we trace its path through period-eccentricity space as it evolves. At each timestep, we add $w\times\mathrm{d}t$ to the probability map, where $w$ is a weight term proportional to the IMF weight, quantifying the number of systems with the same parameters we would expect to occur per $10^6 M_\odot$ of star formation activity. The proportionality term is a weight related to the uniformly sampled kick direction.

The log-likelihood of a given model is then given by

\begin{equation}
    \loge\mathcal{L} = \sum_{n=0}^{14} \loge p(\loge P_n, e_n),
\end{equation}

where $\left(P_n, e_n\right)$ is the period and eccentricity of the $n^\text{th}$ observed BNS system. We can then transform this log-Likelihood into a log-Bayes factor, through

\begin{equation}\label{eq:logbayes}
    \loge\mathcal{K} = \loge\mathcal{L}-\loge\mathcal{L}_\text{Hobbs}.
\end{equation}

where $\loge\mathcal{L}_\text{Hobbs}$ is the log-Likelihood of the Hobbs kick. A $\loge\mathcal{K}>3.4$ ($<-3.4$) is interpreted as very strong evidence for (against) the model relative to the Hobbs kick. The reader is referred to Appendix B of \citet{Jeffreys_1961} for a comprehensive description on the precise numerical values of $\loge\mathcal{K}$. The fourteen BNS systems we use to compute the goodness-of-fit are given in \autoref{tab:observed_BNSes}.

The catalogue of BNS systems we use do not include any Gamma Ray Bursts (GRBs). Even though a short GRB (SGRB) was discovered as the optical counterpart to a gravitational wave event in GW170817 \citep{GW170817_2017}, giving a potential avenue of extra constraints to apply, there are still multiple unknowns which impair our ability to use these as a constraint. For example, \citet{EldridgeStanwayTang_2018} finds that the uncertainty in the SGRB merger rate spans four orders of magnitude, whereas the BNS merger rate from \citet{GWTC-2:Data} only spans 2.86 orders of magnitude. In addition, there are selection effects at play when considering SGRBs. Although the Fermi catalogue has detected $>500$ SGRBs \citep{Gruber_2014,von_Kienlin_2014,Bhat_2016,von_Kienlin_2020}, the majority occur at low redshift, with the peak of the redshift distribution being $z\sim0.9$ \citep{Wanderman_2015}. The dearth of SGRBs at higher redshifts is likely to be due to SGRBs being reasonably faint and thus more likely to be difficult to detect.

\begin{table}
    \centering
    \begin{tabular}{c|c|c|c}
        \hline
        Name & $\log (P / \text{days})$ & $e$ & Ref. \\
        \hline
         J0453+1559 &  0.609808 &  0.113 & \citet{Martinez_2015} \\
         J0737-3039 & -0.991400 &  0.088 & \citet{Kramer_2006} \\
           B1534+12 & -0.375718 &  0.274 & \citet{Fonesca_2014} \\
         J1756-2251 & -0.494850 &  0.181 & \citet{Faulkner_2005} \\
           B1913+16 & -0.490797 &  0.617 & \citet{Hulse_1975} \\
         J1913+1102 & -0.686133 &  0.090 & \citet{Lazarus_2016} \\
         J1757-1854 & -0.735182 &  0.606 & \citet{Cameron_2018} \\
         J1518+4904 &  0.936212 &  0.249 & \citet{Janssen_2008} \\
         J1811-1736 &  1.273672 &  0.828 & \citet{Corongiu_2007} \\
         J1829+2456 &  0.070407 &  0.139 & \citet{Champion_2004} \\
         J1930-1852 &  1.653791 &  0.399 & \citet{Swiggum_2015} \\
         J1753-2240 &  1.134751 &  0.304 & \citet{Keith_2009} \\
         J1411+2551 &  0.417638 &  0.169 & \citet{Martinez_2017} \\
         J1946+2052 & -1.107905 &  0.064 & \citet{Stovall_2018} \\
        \hline
    \end{tabular}
    \caption{The fourteen observed BNS systems we consider. J1930-1852 in particular is an important system to consider: its high period but low eccentricity places it on the periphery of the main `track' of evolution. Thus, our preferred period-eccentricity distribution must be one that replicates this system. This data was originally compiled in \citet{Vigna-Gomez2018OnStars}.}
    \label{tab:observed_BNSes}
\end{table}

\subsubsection{Single-Star Kick Velocities: Cumulative Distribution Functions}\label{sec:cdfs}

We also investigate the goodness-of-fit of the two kick models against observed pulsar velocities on a single-star model grid. We use the dataset from \citet{Willcox_2021}, and filter our list of pulsars in the same way: we remove millisecond pulsars, pulsars in binaries, pulsars in globular clusters, and velocities greater than 2000\kms. This leaves a selection of 81 pulsars forming a reasonably homogeneous dataset in which the velocity of our pulsars is primarily set by the natal kick and previous binary evolution.

Our single star grid spans the same range as \textsc{FineGrid}, and the models originate from three separate sources: BPASS primary models (models of the primary stars in binaries), BPASS secondary models (models of the secondary stars in binaries), and isolated single star models. This allows us to generate model distributions of velocities, which we can convert into probability density functions (PDFs). From here, we may compute the cumulative distribution functions (CDFs) of each model, and of the observed data set, and then perform a KS test. The KS statistic of a given model is defined as the maximum difference between its CDF and the observed CDF, and is thus a measure of the goodness-of-fit of a model to an observed CDF: a low KS statistic indicates a closer fit than a higher one.

\subsubsection{Ultra-Stripped Supernovae: Kick Velocities}\label{sec:kicks_of_ussne}

Our final test involves examining the results of synthetic USSNe. We use the definition of an USSNe from \citet{Tauris_2015}: a supernova which has an envelope mass $M_{\rm env} \leq 0.2M_\odot$ with a compact star companion. For every value of $\alpha$ and $\beta$, we examine the kick velocity produced by the BPASS stellar models. In particular, we only consider models which match the following criteria:

\begin{enumerate}
    \item $0.0 \leq M_{\rm ej} \leq 0.3 M_\odot$, to reflect the low-ejecta mass nature of USSNe (choosing $M_{\rm ej}\leq0.3 M_\odot$ admits the USSNe SN2019dge from \citet{Yao_2020}),
    \item $1.1 \leq M_{\rm rem} \leq 1.8 M_\odot$, from \citet{Tauris_2015}, and
    \item $m_{\rm env} = M-M_{\rm CO} \leq 0.2 M_\odot$, from \citet{Tauris_2015}.
\end{enumerate}

This leaves us with models of secondary stars which will likely undergo a USSNe. For every pair of $\alpha$ and $\beta$, we record the number of kicks between $0 \leq v_{\rm} \leq 30\kms$ that it produces. Since the rate of Type Ic USSNe is estimated to be low -- in the BPASS dataset, it is 1.8 per cent of all CCSNe -- we take the `acceptable region' generated by this test as the region for which at least one such kick is possible.

\section{Results}\label{sec:results}

\subsection{BNS Merger Rates}

\begin{figure*}
    \centering
    \includegraphics[width=\textwidth]{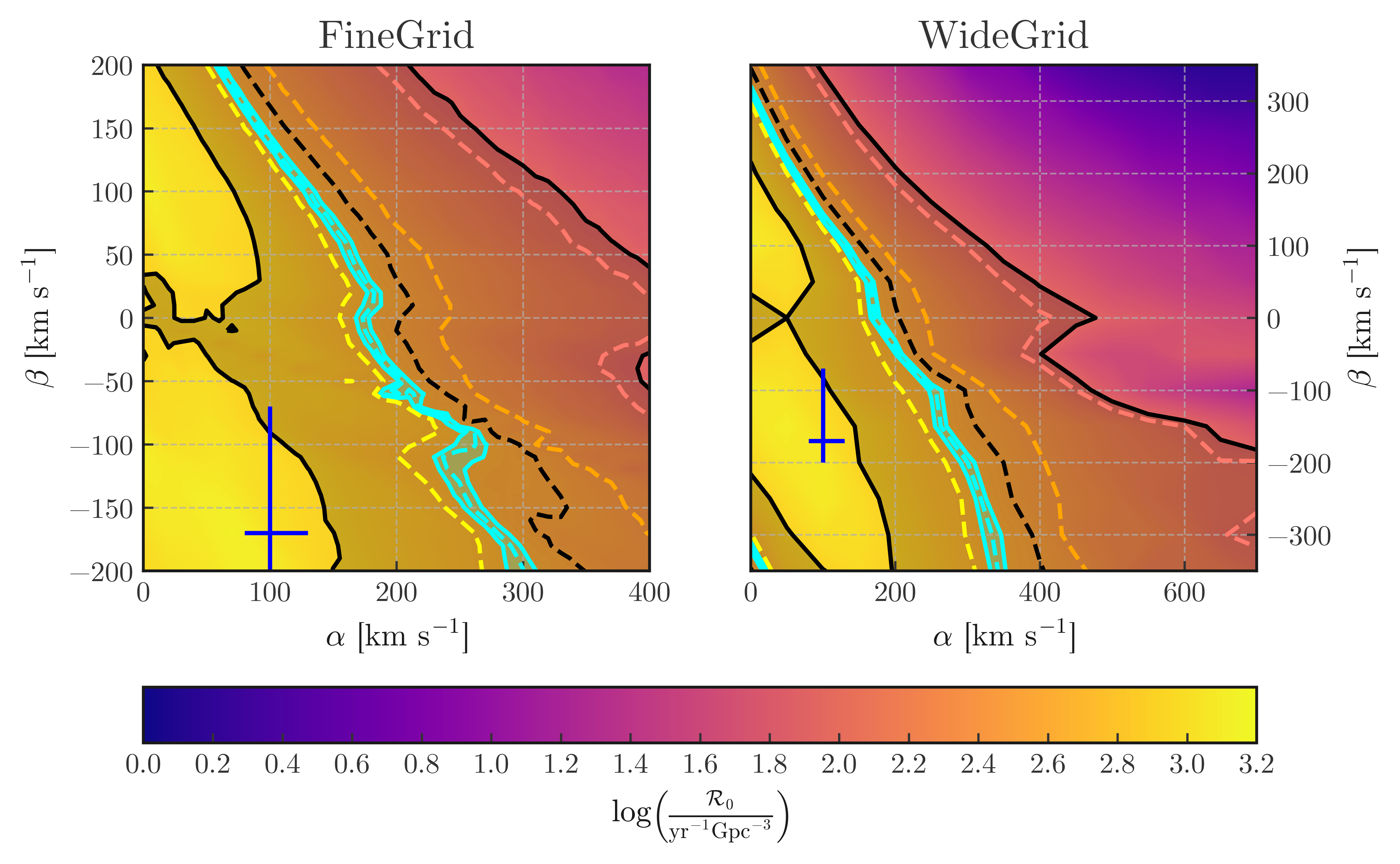}
    \caption{The BNS merger rate distribution on \textsc{FineGrid} and \textsc{WideGrid}. Dashed contours represent reported nominal values, shaded regions represent uncertainties. GWTC-2 data (black contour) is from \citet{GWTC-2:Data}, and GWTC-3 (pink: BGP, yellow: MS, orange: PDB [ind], see \autoref{tab:ligo_rates}) data is from \citet{GWTC-3:Data}. We do not report uncertainties for the GWTC-3 data, as the width of these cover most of the plane. The cyan dashed contour is the theoretical merger rate reported in \citet{EldridgeStanwayTang_2018}. The blue data point is the fiducial value of the Bray kick reported in \citet{Bray2018_RefiningKick}, which does not lie on the cyan contour as we have improved the rejuvenation age calculation in our BPS code, in preparation for the release of new versions of BPASS.}
    \label{fig:braykick_vary_grids}
\end{figure*}

We compute the BNS merger rate on a two dimensional grid of $\alpha$ and $\beta$ combinations. The result is given in \autoref{fig:braykick_vary_grids}. We note that the fiducial value of the Bray kick does not lie on the \citet{EldridgeStanwayTang_2018} contour. This is due to modifications we have made in the BPASS code around the calculation of the rejuvenation age (for details on this, the reader is referred to \citet{Ghodla_2021}, Section 2.1). The BNS merger rates reported by the LVK are given in \autoref{tab:ligo_rates}, and shown against our \textsc{FineGrid} data in \autoref{fig:braykick_vary_grids}.

We find 55 pairs of $\alpha$ and $\beta$ -- 3 per cent of the parameter space -- that agree within the errors of the BNS merger rate from \citet{GWTC-3:Data}. These form a contour on the $\alpha-\beta$ plot in \autoref{fig:braykick_vary_grids}. We term this the `LIGO-Virgo O3a Contour' and discuss this below.

\subsection{Period-Eccentricity Distributions}

\begin{figure*}
    \centering
    \includegraphics[width=\textwidth]{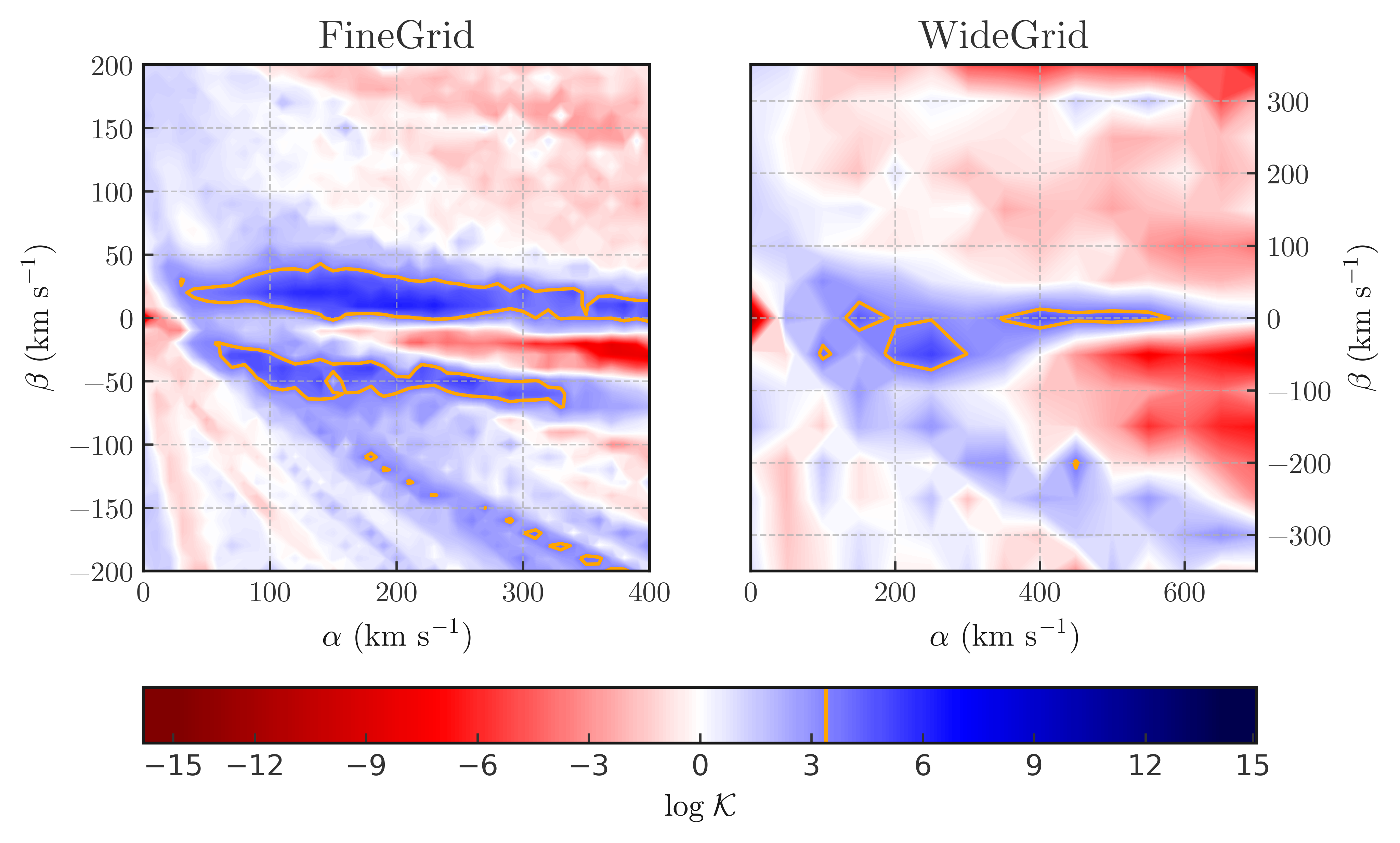}
    \caption{The log-Bayes factor of the period-eccentricity distribution for our grids, measured relative to the Hobbs kick. We note that there is a large range of log-Bayes Factors - including certain pairs of parameters which are more favourable than the Hobbs kick, and certain ones which are preferred less than the Hobbs kick. The orange contour line represents a Bayes factor confidence level: $\loge\mathcal{K}=3.4$.}
    \label{fig:logBayes_plot}
\end{figure*}

In \autoref{fig:55_combinations}, we report the $\loge~\mathcal{K}$ of each system along the LIGO-Virgo O3a Contour alongside their period-eccentricity distributions. We also display the $\loge\mathcal{K}$ values for the all of \textsc{FineGrid} and \textsc{WideGrid} in \autoref{fig:logBayes_plot}. We see better fits towards certain pairs of $\alpha$ and $\beta$ (shown in blue), and less favorable fits towards others (shown in red). A value of $\loge\mathcal{K} > 0$ for a given model implies that the model is more preferable than the Hobbs model, and the exact magnitude of $\loge\mathcal{K}$ indicates how preferable it is.

\autoref{fig:55_combinations} also demonstrates that the structure in the period-eccentricity distribution is sensitive to the natal kick received. For instance, the $\alpha=200\kms, \beta=-50\kms$ kick produces a short strand of higher-period binaries, which is not present in, for example, the $\alpha=90\kms, \beta=140\kms$ distribution. On the other hand, the $\alpha=290\kms,\beta=-180\kms$ distribution has a prominent region in the lower right corner which has no systems. This region is present -- though varies in size -- in 42 of the distributions. The remaining distributions -- for example, $\alpha=90\kms, \beta=140\kms$ -- possess systems that are able to have high-periods and low eccentricity.


\begin{figure*}
    \centering
    \includegraphics[width=\textwidth]{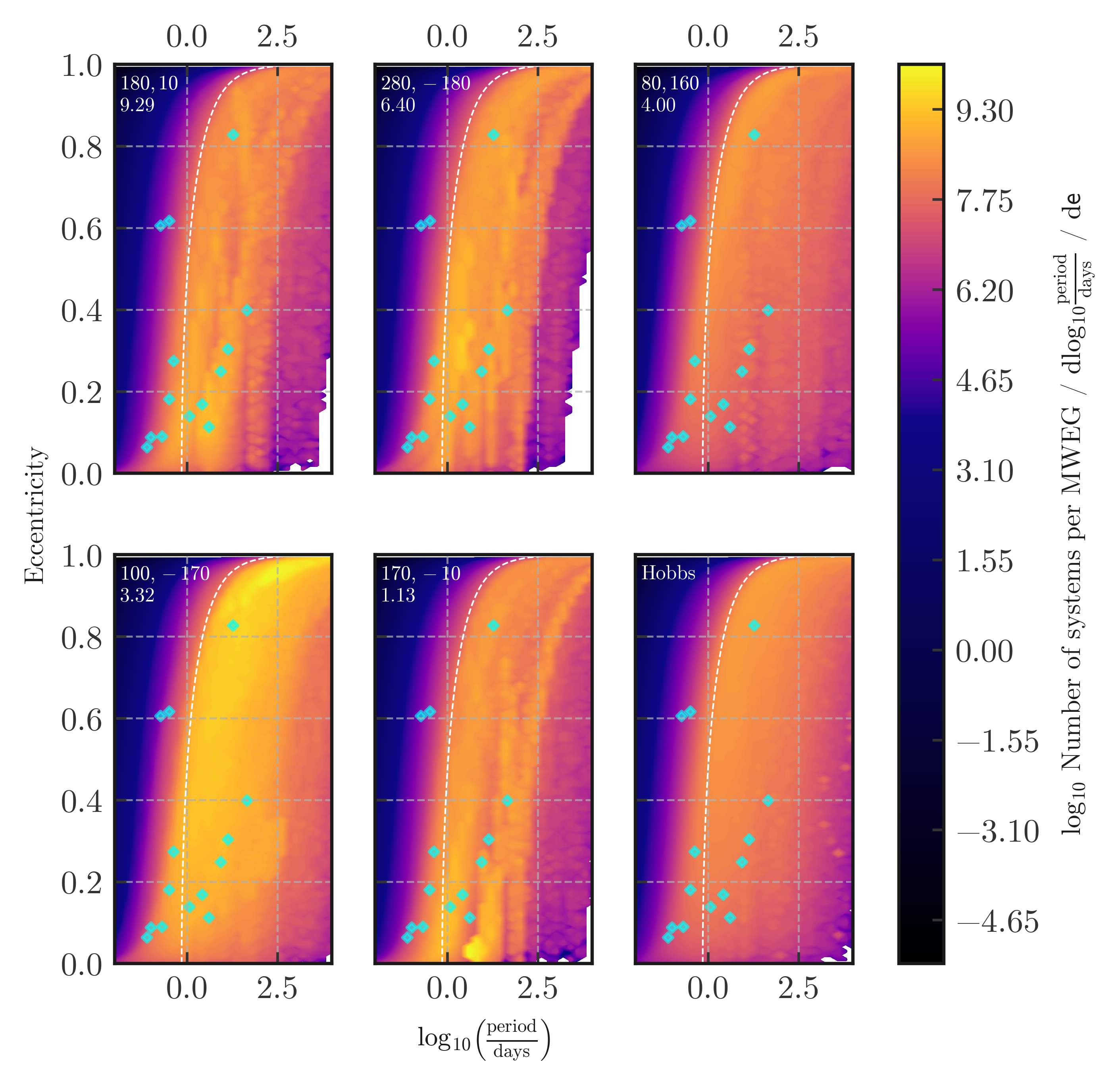}
    \caption{Six period-eccentricity distributions from parameter pairs chosen along the O3 contour. The cyan diamonds are the galactic BNS systems as tabulated in \autoref{tab:observed_BNSes} that are used to compute the log-Bayes factors. The white dashed line is the contour representing a constant-time coalescence of the Hubble time, and is computed using the approximation in \citet{mandel2021accurate}. We select $\alpha=180, \beta=10$ as the parameters with the highest log-Bayes factor from the contour, $\alpha=170, \beta=-10$ as the lowest, and the remainder to illustrate structural differences. Empty regions are spaces where the number of systems per bin is zero. The colourbar has units of `number of systems per MWEG per 0.2 dex per 0.05 eccentricity', as the log-Bayes factor is sensitive to the choice of bin size.}
    \label{fig:4panels}
\end{figure*}

We highlight six distributions in particular. These are shown in \autoref{fig:4panels}. Whilst the general shape of the four distributions are similar, they have some subtle differences. Firstly, and most notably, each plot has a region in the lower right corner - corresponding to low-eccentricity, high-period systems - which has no mergers. This region is smallest for the $\alpha=80, \beta=160$ plot. In general, systems with high periods and low eccentricities have higher coalescence times than other systems. Given that our integrator allows systems to merge for longer than the age of the Universe -- the cut-off time of our integrator is $10^{11}$ years -- we can see that some kicks are able to produce systems that take longer to merge.

We note that of the six distributions plotted in \autoref{fig:4panels}, the location of the densest region varies markedly. For the fiducial Bray kick, the densest region is in the top right region of the space. For $\alpha=170\kms, \beta=-10\kms$, the densest region is around $(\log_{10}P, e)\simeq(1,0)$. We note that $\alpha=170\kms, \beta=-10\kms$ produces a larger kick velocity than the fiducial Bray kick over all positive ejecta-remnant mass ratios. When examining this peak density over the 55 distributions in \autoref{fig:55_combinations}, we find that kicks with a higher average velocity produce peaks in the lower left region of the plot -- indicating lower period, circular binaries.

\subsection{Single-Star Kick Velocities}

\begin{figure*}
    \centering
    \includegraphics[width=\textwidth]{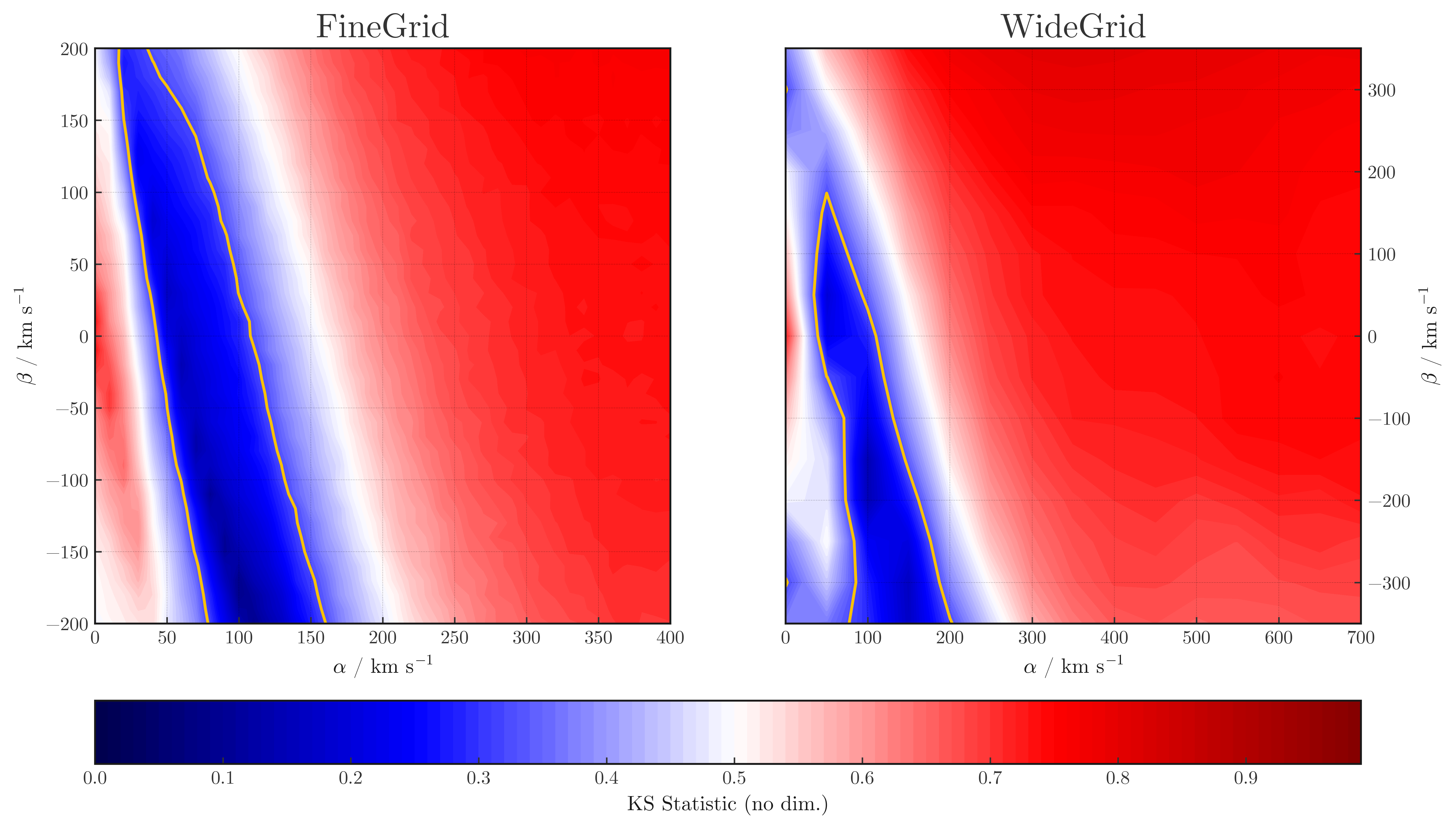}
    \caption{The KS statistic for each combination of $\alpha$ and $\beta$ across a single-star version of \textsc{FineGrid} and \textsc{WideGrid}. The values of $\alpha$ and $\beta$ that have the lowest KS statistic are $\alpha=100, \beta=-180$ on \textsc{FineGrid}, and $\alpha=100, \beta=-150$ on \textsc{WideGrid}. The orange contour line is the KS statistic for the Hobbs kick, which has a KS statistic of 0.33.}
    \label{fig:ks_test}
\end{figure*}

Applying the methods outlined in Section~\ref{sec:cdfs}, we see in \autoref{fig:ks_test} that the most favorable model is $\alpha=90\kms, \beta=-170\kms$ -- almost replicating the earlier work of \citet{Bray2018_RefiningKick}. This is expected, as we are effectively following in the steps of \citet{Bray2018_RefiningKick}, albeit with a different compiled dataset. In general, we find a valley of well-fitting models, with the best fits in the region roughly bounded by the triangle with corners at $\alpha=90\kms, \beta=-150\kms$, $\alpha=90\kms, \beta=-200\kms$, and $\alpha=120\kms, \beta=-200\kms$. 

\begin{figure*}
    \centering
    \includegraphics[width=\textwidth]{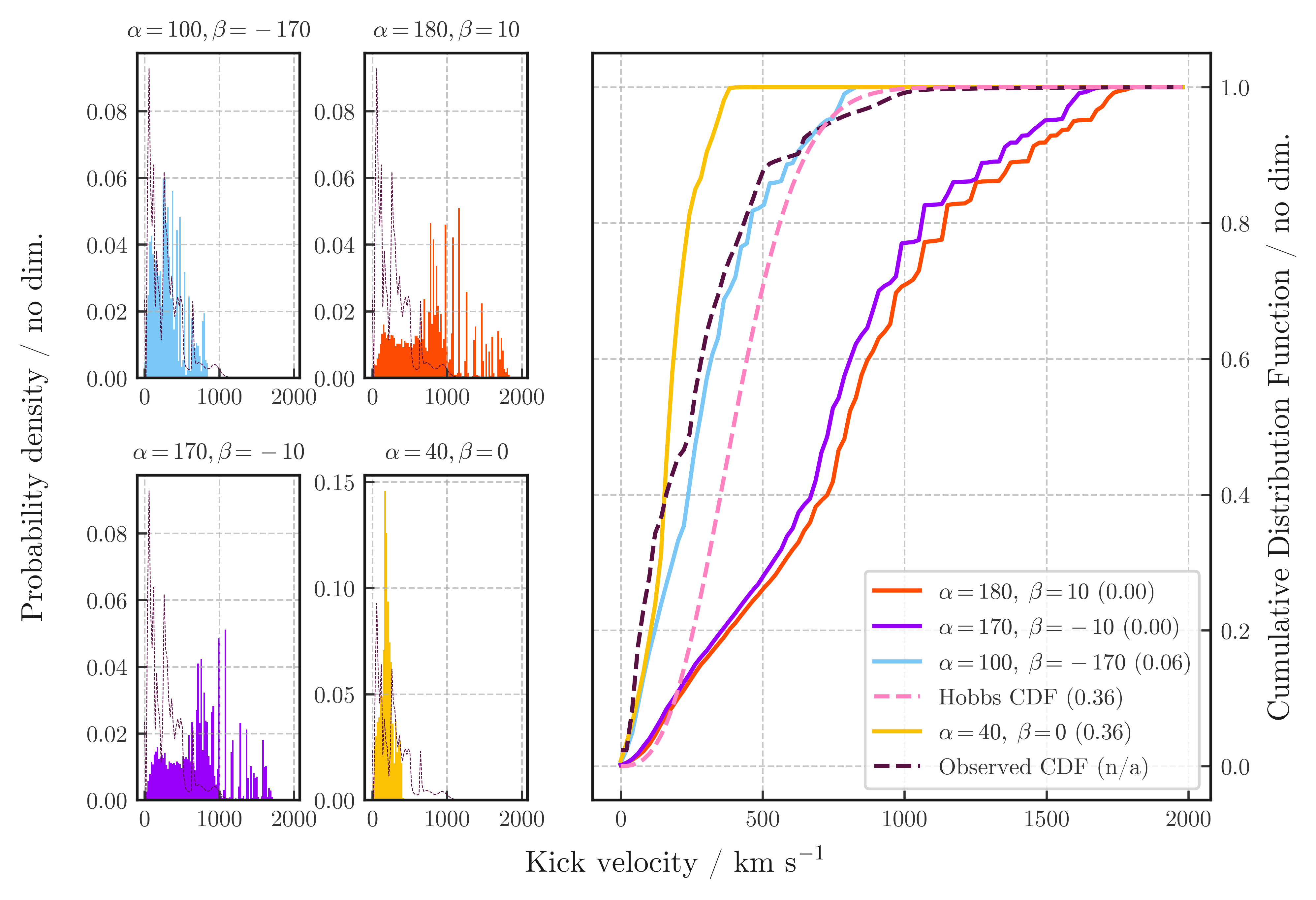}
    \caption{The PDF for the single-star velocity distribution for four combinations of $\alpha$ and $\beta$. We select $\alpha=100\kms,~\beta=-170\kms$ as the fiducial Bray kick from \citet{Bray2018_RefiningKick}, $\alpha=180\kms,~\beta=10\kms$ as the combination of parameters with the highest $\loge\mathcal{K}$ from the LIGO-Virgo O3a contour, $\alpha=170\kms,~\beta=-10\kms$ as the combination of parameters with the lowest $\loge\mathcal{K}$, and $\alpha=40\kms,~\beta=0\kms$ to illustrate a system with a low $\alpha$ and zero $\beta$. The `empty' bins in the $\alpha=180\kms,~\beta=10\kms$ and $\alpha=170\kms,~\beta=-10\kms$ PDFs have contents on the order of $10^{-3}$. The dashed purple line is the PDF corresponding to the observed PDFs from \citet{Willcox_2021}. In the right panel, the pink dashed line is the CDF corresponding to the Hobbs kick, which has a KS statistic of 0.32. Also shown in the rightmost panel is the CDF for each model, where the colours of the lines correspond to the colours of the relevant histograms. The number in parentheses in the legend is the KS statistic of the given model, which does not apply to the observed CDF as that is the reference model we measure against.}
    \label{fig:individual_pdfs}
\end{figure*}

We include the PDFs of four models, shown in \autoref{fig:individual_pdfs}. The four models we choose are the fiducial Bray kick, the kicks with the highest and lowest $\loge\mathcal{K}$ values, and a kick with a low $\alpha$ and zero $\beta$. We show them relative to the observed PDF of the single-star pulsars compiled in \citet{Willcox_2021}. Even the model with the lowest KS statistic -- $\alpha=90\kms, \beta=-170\kms$ -- has difficulty reproducing the observed PDF, though the fit to the CDF is reasonable. This is likely due to the sensitivity of the PDF to the chosen bin size, whereas the CDF is insensitive to the (essentially arbitrary) selection of bin size. \autoref{fig:individual_pdfs} also shows the CDF of each of the four models against the observed CDF. The model which has the highest $\loge\mathcal{K}$ has a KS statistic of 0.61, which indicates it diverges quite strongly from the observed CDF. Indeed, we see that although it does produce kicks $\lesssim500\kms$, it underpredicts the probability of those kicks quite markedly.

\subsection{Production of Ultra-Stripped Supernovae}

We find a reasonably narrow region of the parameter space permits kick velocities consistent with USSNe. The region is broadly consistent with the analysis that the production of USSNe requires low $\beta$ values. The region is broadest in $\beta$ towards the right edge of the plot, coinciding with high $\alpha$. This is consistent with theory: if we consider an ejecta mass of $0.1M_\odot$ resulting in a Neutron star of mass $1.4M_\odot$, both $\alpha=30\kms, \beta=10\kms$ and $\alpha=100\kms, \beta\simeq-20\kms$ would produce weak kicks of about $10\kms$. To constrain the kick to have $\beta\simeq0$, we require the other three constraints.

\subsection{Combining the Constraints}\label{sec:all_constraints}

The four major constraints that we have established -- the event rate in \autoref{fig:braykick_vary_grids}, the log-Bayes factor in \autoref{fig:logBayes_plot}, the KS statistic in \autoref{fig:ks_test}, and the space which permits USSNe -- enable us to define heuristics that admit only a subset of our parameter space as acceptable.

We set the following constraints:

\begin{enumerate}
    \item The event rate of the model must lie along the LIGO-Virgo O3a Contour, or at worst within the uncertainty bounds,
    \item The log-Bayes factor must be above 4.61, which represents decisive evidence for the model against the Hobbs kick,
    \item The KS statistic must, conservatively, be below 0.5, and
    \item The parameter space region must be capable of producing USSNe.
\end{enumerate}

We note that the KS statistic constraint is essentially arbitrarily set. We select 0.5 as the point where the data fits the observed CDF better than it diverges from it. Varying the KS statistic threshold results in between 0 models (if a threshold of 0.27 or fewer is selected) and 31 models (if a threshold of 0.74 or higher is selected).

\begin{figure*}
    \centering
    \includegraphics[width=2\columnwidth]{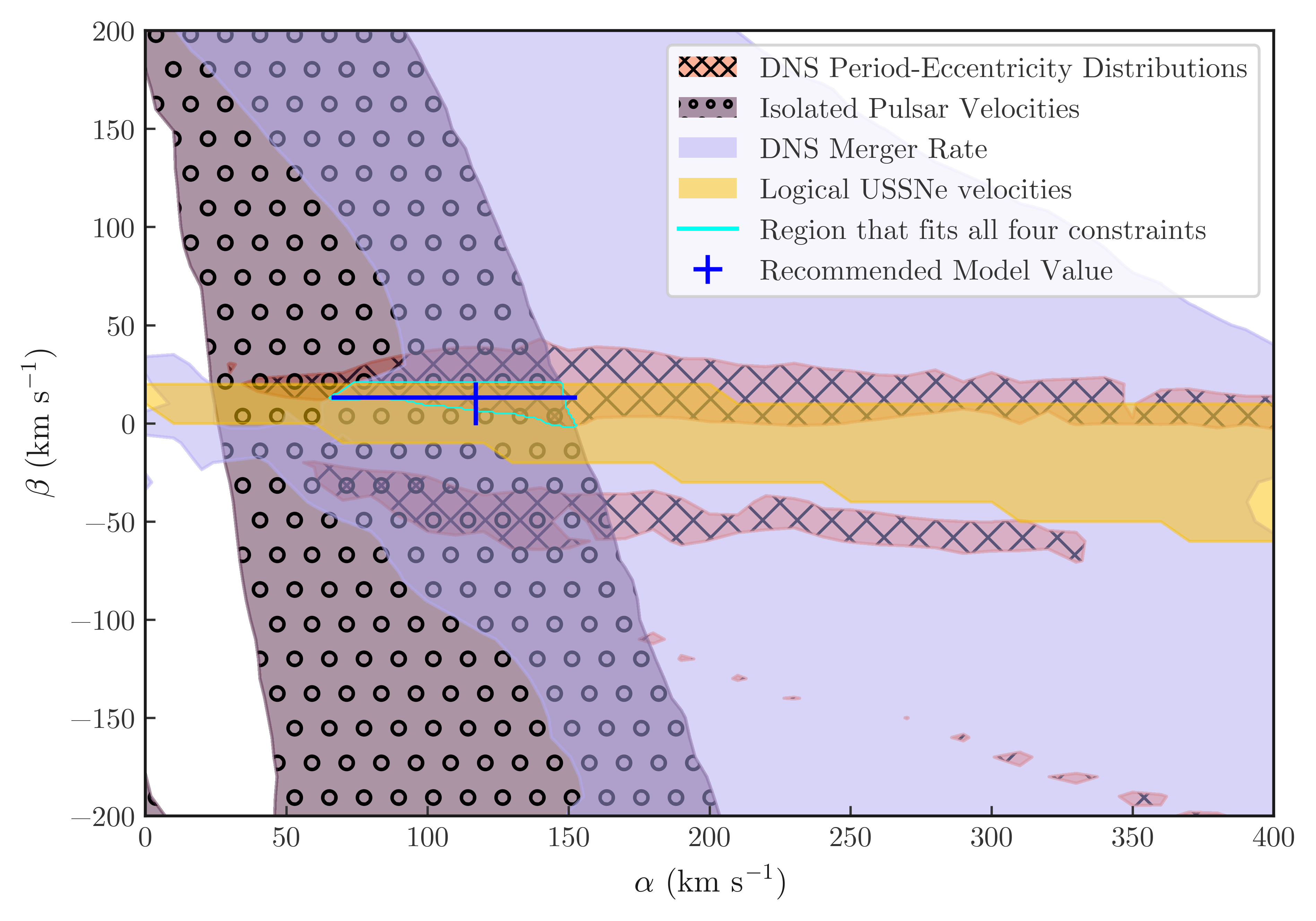}
    \caption{The regions of \textsc{FineGrid} which fit individual parts of our constraints. The cyan contour demarcates the regions of the space which fit all four criteria. The values extracted from within the cyan contour are tabulated in \autoref{tab:all_matching_fits}. The cyan contour is computed on a finer grid than the simulations are run on. As such, the blue cross, which is centered on the ``center of mass'' of the cyan contour and lies at $\alpha=117^{+36}_{-51}\kms, \beta=13^{+8}_{-14}\kms$, is likely an overfit. To accommodate for this, the value we recommend is adjusted to lie on the nearest off-grid data point (i.e., the nearest multiple of half of the bin width, or $5\kms$) in \textsc{FineGrid}, with error bars adjusted in the same way (broadened, where necessary).}
    \label{fig:entire_parameter_space}
\end{figure*}

We can overplot the relevant regions from \autoref{fig:braykick_vary_grids}, \autoref{fig:logBayes_plot}, and \autoref{fig:ks_test} into a single plot -- \autoref{fig:entire_parameter_space} -- and use these to identify which regions of our parameter space fit our four constraints. Applying the four constraints to our dataset, we find 1 per cent of our parameter space fits all four constraints. These are tabulated in \autoref{tab:all_matching_fits}, along with their corresponding KS statistics, $\loge\mathcal{K}$ values, $\log~\mathcal{R}_0$ values, and redshifts that correspond to the peak in the BNS merger rate distribution.

\begin{table*}
    \centering
    \begin{tabular}{c|c|c|c|c|c|c|c|c|c|c|c|c}
    \hline
    $\alpha$ & $\beta$ & $\log~\mathcal{K}$ & KS stat & $\log~\mathcal{R}_0$ & $z_\text{peak}$ &      & $\alpha$ & $\beta$ & $\log~\mathcal{K}$ & KS stat & $\log~\mathcal{R}_0$ & $z_\text{peak}$ \\
    \hline
    80   & 20   & 4.29 & 0.24 & 2.90 & 0.00 &      & 90   & 20   & 4.66 & 0.27 & 2.88 & 0.00 \\
    100  & 10   & 3.47 & 0.30 & 2.83 & 0.00 &      & 100  & 20   & 5.27 & 0.32 & 2.88 & 0.00 \\
    110  & 10   & 3.71 & 0.35 & 2.82 & 0.00 &      & 110  & 20   & 5.34 & 0.37 & 2.85 & 0.00 \\
    120  & 10   & 4.41 & 0.40 & 2.80 & 0.00 &      & 120  & 20   & 5.72 & 0.41 & 2.82 & 0.02 \\
    130  & 10   & 4.62 & 0.44 & 2.76 & 0.18 &      & 130  & 20   & 5.81 & 0.45 & 2.79 & 0.00 \\
    140  & 10   & 4.65 & 0.47 & 2.73 & 0.26 &      & 140  & 20   & 5.84 & 0.48 & 2.75 & 0.06 \\
    150  & 0    & 4.43 & 0.49 & 2.69 & 0.21 &      &   -  &   -  &   -  &   -  &   -  &   -  \\
    \hline
    \end{tabular}
    \caption{All 13 models that fit the three constraints in \autoref{sec:all_constraints} - that is, $\loge\mathcal{K}\geq3.4$, KS statistic $\leq$ 0.5, capability of producing USSNe, and $\log~\mathcal{R}_0$ on the LIGO-Virgo O3a Contour (or within its uncertainty bound). Uncertainties have been omitted for brevity, and all values have been rounded to 2 decimal places. We also report $z_\text{peak}$, which is the redshift $z$ at which the BNS merger rate distribution peaks, for each model.}
    \label{tab:all_matching_fits}
\end{table*}

\begin{figure*}
    \centering
    \includegraphics[width=\textwidth]{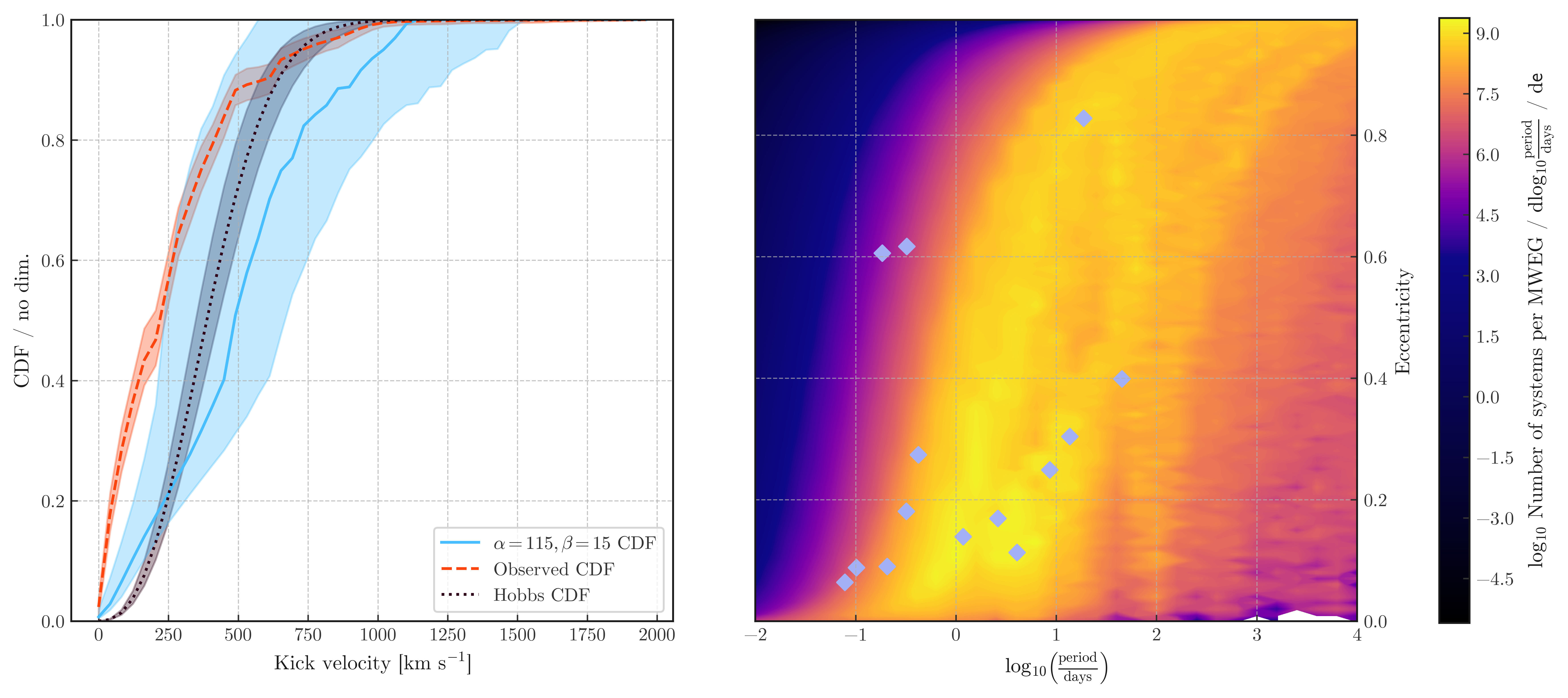}
    \caption{The CDF for a population of isolated pulsars (left) and the period-eccentricity distribution (right) corresponding to $\alpha=115\kms, \beta=15\kms$. The red dashed line on the CDF is the observed CDF from \citet{Willcox_2021}. The shaded blue region represents the CDFs within the uncertainty bands of our best model value, and were computed by computing the CDFs of every combination of $\alpha$ and $\beta$ within the range, in steps of 5\kms. The shaded red region is the uncertainty bounds on the observed CDF, generated by simulating 2,000 samples of the PDF, 10,000 times, and plotting the extent of the obtained CDFs. The shaded black region is the uncertainty on the Hobbs CDF, obtained with $\sigma_\text{max} = 261\kms, \sigma_\text{min} = 239\kms$, from \citet{VerbuntIgoshevCator}.}
    \label{fig:best_model_params}
\end{figure*}

We take the `best parameters' as the center of mass of the cyan region, with uncertainty bounds covering said region. These parameters are $\alpha=117^{+36}_{-51}\kms, \beta=13^{+8}_{-14}\kms$, which is likely an overfit due to the finer grid used to calculate the value (see \autoref{fig:entire_parameter_space}), and as such the best parameters we report are $\alpha=115^{+40}_{-55}\kms, \beta=15^{+10}_{-15}\kms$. The CDF and period-eccentricity distribution for this pair of parameters are shown in \autoref{fig:best_model_params}. We can see it is an adequate match to the CDF and period-eccentricity distributions by eye, and it has a merger rate of $\log_{10}\mathcal{R}_0 = 2.84490\pm 0.00009$, where the uncertainty is assumed to be Poissonian.

Finally, we note that the precise determination of the contour lines is accomplished via a marching squares algorithm in \textsc{matplotlib}. This has the added requirement of using linear interpolation to place the precise locations of the contours on the grid. Unfortunately, in current versions of \textsc{matplotlib} this is not possible to adjust. However as our recommended value lies off-grid and is likely an overfit, the adjustment back to $\alpha=115^{+40}_{-55}\kms, \beta=15^{+10}_{-15}\kms$ accounts for this uncertainty.

\section{Discussion \& Conclusions}\label{sec:discussion}

We found a selection of 13 different pairs of parameters that fit our four constraints. For values of $\alpha\geq 100\kms$, we found that low magnitude, but non-zero, values of $\beta$ replicated the BNS merger rate from \citet{GWTC-3:Data} and replicating the observed BNS systems compiled in \citet{Vigna-Gomez2018OnStars}. A true `best' set of parameters likely lie within this subsample of our parameter space.

PSR J1930-1852 \citep{Swiggum_2015} is the widest known binary and, per its authors, `its unique spin and orbital parameters challenge
models that describe BNS formation'. Whilst our work, unlike other authors such as \citet{Tauris_2017}, did not preferentially weight our binaries -- that is, we did not prefer models that replicate extreme parameters such as PSR J1930-1852 -- we do note that a number, but not all, of our period-eccentricity distributions had non-zero probability of reproducing this pulsar. Thus, certain kick prescriptions will be able to replicate its unique properties.

\begin{figure*}
    \centering
    \includegraphics[width=\textwidth]{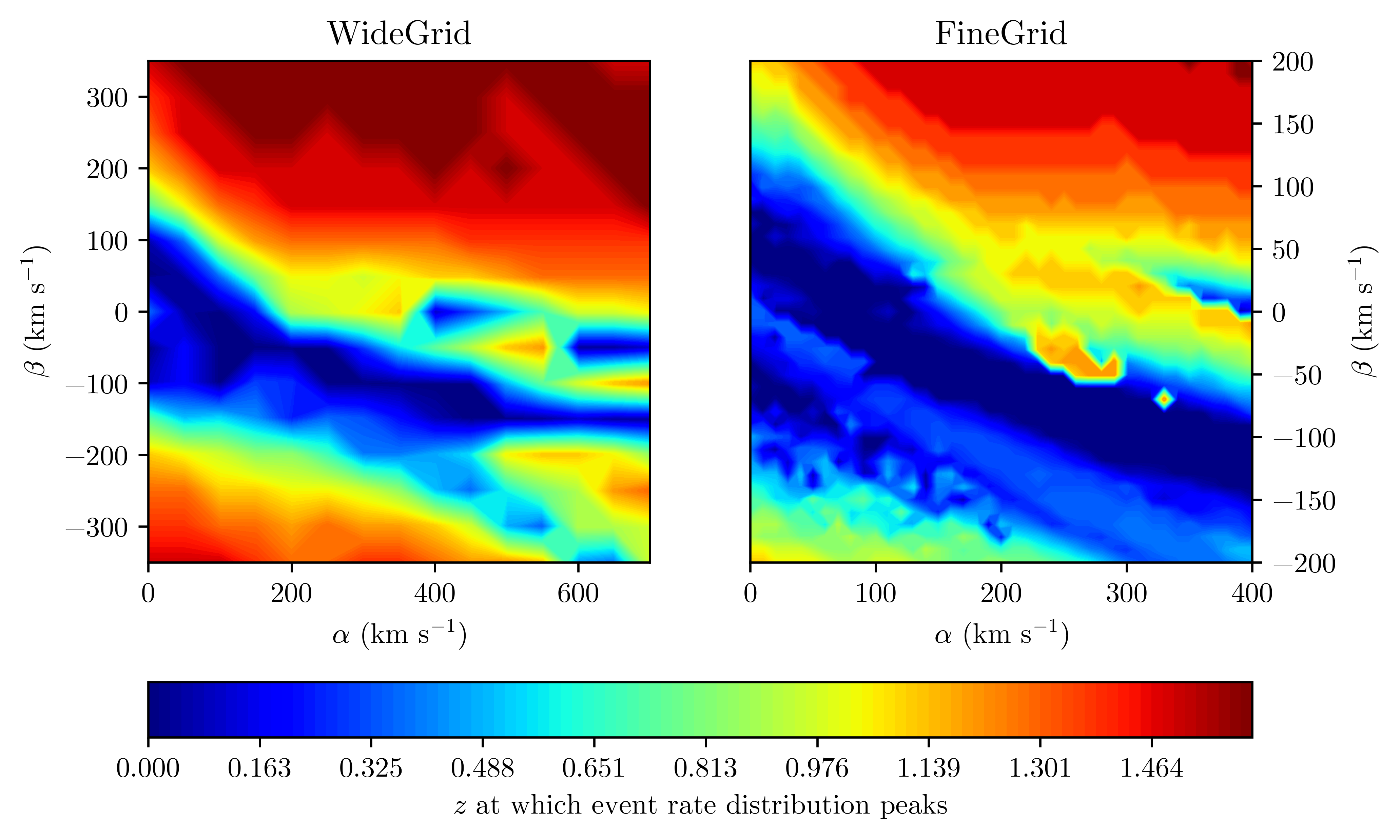}
    \caption{The location in redshift $z$ at which the event rate distribution peaks, as a function of $\alpha$ and $\beta$.}
    \label{fig:shifting_peak}
\end{figure*}

Whilst the current observed BNS merger rate is limited, and has wide uncertainty bands, further observations will allow the distribution of the merger rate over redshift to become more refined. Since the peak of the event rate is dependent on the DTD, this distribution is also sensitive to the natal kick and thus can be used as a further constraint on the Bray kick parameters In \autoref{fig:shifting_peak} we display the location in redshift of this peak for our \textsc{FineGrid} and \textsc{WideGrid} model sets. We find that the peak redshift across both grids corresponds to $z<1.5$, with a very clear and defined valley of the present day. Across the 13 pairs of parameters that fulfil all of our constraints, we also find that the peak corresponds to $z\leq0.26$, and the best model $\alpha=115\kms, \beta=15\kms$ peaks in the present-day, at $z=0$. 

The primary limitation with our $z_\text{peak}$ analysis is a lack of observational data to compare against. Although factors like the galactic stellar formation rate from \citet{Madau2014CosmicHistory} are derived as fits to empirical measurements, given that a redshift of $z=2$ (higher than the maximum of the grids) corresponds to a lookback time of $t_\text{L}\simeq 10$ Gyr, we simply do not have direct observational evidence to use this as a constraint to kick models. However, indirect evidence is present in the enrichment history of r-process elements in the Universe. \citet{vandeVoort}, for example, finds that natal kicks influence the number of r-process enhanced stars in the Universe. Their analysis indicates that for kicks with high velocities, which cause the infant neutron stars to be ejected from the galaxy, r-process elements are no longer able to accrete onto the star-forming galaxy and therefore can not contribute the elements to new generations of stars.

We have also highlighted the caveat that $\alpha=115^{+40}_{-55}\kms, \beta=15^{+10}_{-15}\kms$ should not be interpreted as `the correct choice of parameters' but rather `the mean of the set of parameters which replicates the four observables we test against'. It is also pertinent to note that these values are calibrated against the BPASS version 2.2.1 models (and should not change significantly when used with future stellar models from the same stellar evolution code). Using these values with other BPS codes should be approached with caution, and readers using other codes should calibrate their choice of parameters against their BPS codes in the same way that we have.

As part of the calibration and selection of binary parameters, we simulated 1,000 kicks per binary. Testing showed a slight, but not unexpected, dependence on number of simulated kicks. We found that adjusting the number of kicks simulated by a factor of ten resulted in a shift of $\loge\mathcal{K}$ by approximately one.

Similar work in this field has been done for binary Black-Hole (BBH) mergers by e.g., \citet{Mapelli_2018, Wiktorowicz_2019, Giacobbo_2020, Santoliquido_2021}, however few authors have explored the parameter space in the way that we have. The most similar studies have varied the properties of the progenitor population -- they have adjusted the IMF, or the stellar formation rate, or DTD, and assumed a kick prescription with given free parameters to be true. With this study, those population properties have been affixed, and we have varied the observed parameters we aim to replicate, to see which set of parameters replicates all four observables best.

Authors such as \citet{Fryer_1998, Arzoumanian_2002, Bombaci_2004} suggest that the kick distribution should ostensibly contain two peaks. In testing, we found no combination of $\alpha$ and $\beta$ for the Bray kick produced an obviously bimodal single-star velocity distribution, though we do observe interesting structure in the resultant PDFs. \autoref{fig:best_model_params} clearly shows several changes of gradient in the CDF, which can be interpreted as several ranges of kick behaviour. This behaviour arises because the Bray kick is linked to the ejecta mass distribution. Thus the behaviour of low/high kick populations could be due to the results of stellar evolution causing low/high ejecta masses for CCSNe that form neutron stars. 

The reason we did not include a bimodal Hobbs kick in this study is twofold. Firstly, the unimodal Hobbs kick has been extensively discussed in literature \citep[see, among others,][]{Nakar_2007, Belczynski_2008, Lorimer_2008, Dominik_2012, Bray2016_ProposingKick, Eldridge_2017, Bray2018_RefiningKick} and thus was a good starting point for our investigation. Secondly. as a point of consideration, we computed the log-Likelihood of reproducing our BNS catalog against a weighted model from \citet{VerbuntIgoshevCator}, taking $w=0.42,\,\sigma_1=75\kms,\,\sigma_2=316\kms$. We found that $\loge\mathcal{K} = 0.8$. Thus, if this value were to be used as the reference model for the study, it would shift our log-Likelihoods by approximately one order of magnitude and would only admit an extra handful of systems.

In the future, an additional prior can be used on our dataset. The natural interpretation of the Bray kick expression is that $\beta$ is the `limiting value' that the kick velocity should approach as the ejecta mass gets smaller (as in the case of USSNe and ECSNe). In events such as the direct collapse of a star to a black hole in a failed SNe, where the ejecta mass is zero, $\beta$ would identically be the kick velocity. Thus, in order for conservation of momentum to be respected for low-ejecta mass kicks, we expect $\left|\beta\right|>0$ -- i.e., we expect the magnitude of $\beta$ to be low, but most likely non-zero. This prior was not imposed on our parameter space -- the $\loge\mathcal{K}$ and USSNe constraints independently restricted $\beta$ to be >0. It is pertinent to note, as well, that even in the limit as the ejecta mass goes to zero, a value of $\beta<0$ would imply that the velocity vector is anti-aligned with the chosen direction of the kick. $\beta<0$ also allows for a zero kick velocity when the ejecta mass is non-zero. At this stage, this scenario cannot be ruled out, thus we cannot yet include this constraint as a prior.

In addition, we found that the mean pair of parameters -- the center of the cyan region in \autoref{fig:entire_parameter_space} -- corresponded to $\alpha=115^{+40}_{-55}\kms, \beta=15^{+10}_{-15}\kms$. Although this work does not purport to claim a single `best' pair of parameters, these ones can be interpreted as the pair that optimises each of the individual tests that we have performed.  We note that this value lies off-grid, and hence in \autoref{fig:best_model_params} we generated a bespoke data file for this set of two parameters.

In sum, the key conclusions we draw are:

\begin{enumerate}
    \item Providing multiple constraints allows for the parameter space of possible kick configurations to be drastically reduced.
    \item Admissibility of kick velocities of $\lesssim 30\kms$ for USSNe is one of the tightest constraints we can provide.
    \item The peak of the merger rate distribution is likely to be the present-day, and this is a further constraint we can apply to the dataset when more information is available.
    \item The center of mass of the permitted region of the Bray parameter space is $\alpha=115^{+40}_{-55}\kms, \beta=15^{+10}_{-15}\kms$, although caution should be exercised before using this value without calibration.
\end{enumerate}

The analysis presented here is an analysis pipeline to refine the properties of kick models against four different observables. In principle, it can be applied to any kick model to validate the model. Our statistical analysis provides a simple way to validate kick models: it can be used to rule out either magnitudes of kick velocities or entire physical kick prescriptions.

\section*{Acknowledgements}

SMR acknowledges support from The University of Auckland, and would like to thank Ilya Mandel and Team COMPAS for productive discussions. JJE, MMB, and HFS acknowledge support by the University of Auckland and funding from the Royal Society Te Ap\=arangi of New Zealand Marsden Grant Scheme. RW is supported by the Australian Research Council Centre of Excellence for Gravitational Wave Discovery (OzGrav), through project number CE170100004. The authors wish to thank the NeSI Pan Cluster. New Zealand’s national facilities are provided by the NZ eScience Infrastructure and funded jointly by NeSI's collaborator institutions and through the Ministry of Business, Innovation \& Employment's Research Infrastructure programme. URL: https://www.nesi.org.nz. The authors also thank the anonymous referee for constructive comments on the manuscript.

\section*{Data Availability}\label{sec:dataavailability}

Our analysis code is available on GitHub\footnote{\url{https://github.com/krytic/takahe}}, and our bespoke BPASS outputs are available upon request, though we caution that the larger grids comprise some 300 GB of data when uncompressed.


\bibliographystyle{mnras}
\bibliography{Richards_2022}


\appendix

\section{Plots along the LIGO-Virgo O3a contour}

In this section, we provide plots of the 55 models that lie along the LIGO-Virgo O3a contour. The colour axis is the log of the number of systems per Milky-Way Equivalent Galaxy (MWEG), per period bin and per eccentricity bin. Bins were chosen such that ${\rm d}{e}=0.01$ and ${\rm d}\log P=0.2$. Naturally, eccentricity bins span from $e=0$ to $e=1$, and log-period bins span from $\log P=-2$ to $\log P=4$. Moreover, we assume that there is 1 MWEG per $(4.4~\text{Mpc})^3$, an assumption from \citet{Kasen_2017}.

\begin{figure*}
    \centering
    \includegraphics[height=0.9\textheight]{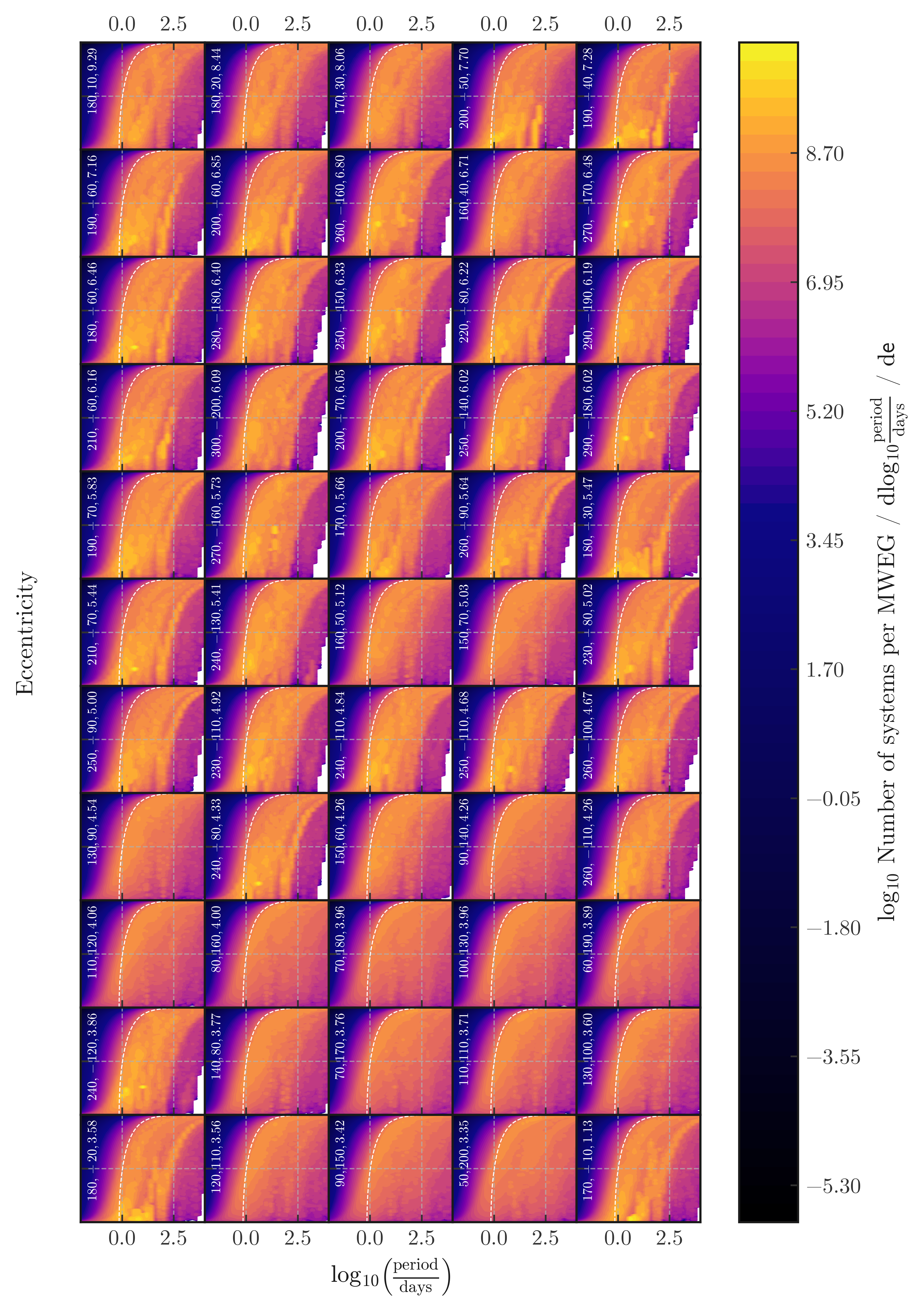}
    \caption{The period-eccentricity distributions for the 55 combinations of $\alpha, \beta$ along the LIGO-Virgo O3a contour. The white dashed curve is a line of constant merger time of $t=t_\text{H}\approx 13.9$ Gyr, for two 1.4 $M_\odot$ remnants. We assume one Milky-Way equivalent galaxy per (4.4 Mpc)$^3$, from \citet{Kasen_2017}. We use the approximation from \citet{mandel2021accurate} for our calculation of the constant merger time contour. The text in the top left corner of each plot denotes the value of $\alpha$ and $\beta$ that generated that plot, alongside its $\loge~\mathcal{K}$ value. The diagrams are sorted such that the highest $\loge~\mathcal{K}$ is in the top left corner, and the lowest $\loge~\mathcal{K}$ value is in the bottom right.}
    \label{fig:55_combinations}
\end{figure*}


\bsp	
\label{lastpage}
\end{document}